\documentstyle[12pt]{article}

\makeatletter
\def\vereq#1#2{\lower3pt\vbox{\baselineskip1.5pt \lineskip1.5pt
\ialign{$\m@th#1\hfill##\hfil$\crcr#2\crcr\sim\crcr}}}
\makeatother

\def\lesssim{\mathrel{\mathpalette\vereq<}}
\def\gtrsim{\mathrel{\mathpalette\vereq>}}

\begin{document}

\begin{titlepage}
\begin{center}
\today     \hfill    LBNL-39422 \\
~{} \hfill UCB-PTH-96/40  \\

\vskip .25in

{\large \bf TeV Right-handed Neutrinos and \\
the Flavor-symmetry-improved
Seesaw Mechanism}\footnote{This 
work was supported in part by the Director, Office of 
Energy Research, Office of High Energy and Nuclear Physics, Division of 
High Energy Physics of the U.S. Department of Energy under Contract 
DE-AC03-76SF00098.  HM was also supported in part by the National 
Science Foundation under grant PHY-95-14797 and by the Alfred P. Sloan 
Foundation.}

\vskip 0.3in

Christopher D. Carone$^1$ and Hitoshi Murayama$^{1,2}$

\vskip 0.1in

{{}$^1$ \em Theoretical Physics Group\\
     Earnest Orlando Lawrence Berkeley National Laboratory\\
     University of California, Berkeley, California 94720}

\vskip 0.1in

{{}$^2$ \em Department of Physics\\
     University of California, Berkeley, California 94720}

\end{center}

\vskip .1in

\begin{abstract}
Horizontal flavor symmetries can drastically suppress Dirac neutrino
masses well below those of the corresponding charged leptons.  We show
that models can be constructed where the light neutrino mass eigenvalues
are small enough to give the MSW solution to the solar neutrino problem,
with a right-handed neutrino scale no larger than a TeV.  We present a
model of this type where the right-handed neutrino scale is generated by
the spontaneous breakdown of gauged U(1)$_{B-L}$, in a radiative
breaking scenario driven by the right-handed neutrino Yukawa couplings.
The model allows for a solution to the $\mu$ problem, and predicts the
existence of a $Z'$ boson within the reach of the LHC or the Tevatron.
\end{abstract}

\end{titlepage}
\renewcommand{\thepage}{\roman{page}}
\setcounter{page}{2}
\mbox{ }

\vskip 1in

\begin{center}
{\bf Disclaimer}
\end{center}

\vskip .2in

\begin{scriptsize}
\begin{quotation}
This document was prepared as an account of work sponsored by the United
States Government. While this document is believed to contain correct 
information, neither the United States Government nor any agency
thereof, nor The Regents of the University of California, nor any of their
employees, makes any warranty, express or implied, or assumes any legal
liability or responsibility for the accuracy, completeness, or usefulness
of any information, apparatus, product, or process disclosed, or represents
that its use would not infringe privately owned rights.  Reference herein
to any specific commercial products process, or service by its trade name,
trademark, manufacturer, or otherwise, does not necessarily constitute or
imply its endorsement, recommendation, or favoring by the United States
Government or any agency thereof, or The Regents of the University of
California.  The views and opinions of authors expressed herein do not
necessarily state or reflect those of the United States Government or any
agency thereof, or The Regents of the University of California.
\end{quotation}
\end{scriptsize}

\vskip 2in

\begin{center}
\begin{small}
{\it Lawrence Berkeley Laboratory is an equal opportunity employer.}
\end{small}
\end{center}

\newpage
\renewcommand{\thepage}{\arabic{page}}
\setcounter{page}{1}
\section{Introduction} \label{sec:intro}

The seesaw mechanism \cite{seesaw} has been proposed as a natural 
explanation for the lightness of the three known neutrino species.  In 
models with right-handed neutrinos, the `seesaw' refers to the 
widely disparate eigenvalues of the neutrino mass matrix
\begin{equation}
\left(\begin{array}{cc}    \sim 0  &  M_{LR} \\ 
              (M_{LR})^T  &  M_{RR}  \end{array}\right)
\,\,\, ,
\end{equation}
where $M_{LR}$ are the entries generated through ordinary electroweak
Higgs couplings, and $M_{RR}$ are the Majorana masses for the
right-handed states.   Since the the right-handed neutrinos are
singlets under the standard model gauge group, they can develop
masses that are much larger than the electroweak scale.  The
result is immediate in the case of one neutrino flavor: the eigenvalues 
of the two-by-two matrix above are of order $M_{RR}$ and $M_{LR}^2/M_{RR}$, 
assuming $M_{RR} \gg M_{LR}$.

This result is significant in light of the experimental indications that 
the light neutrino mass eigenvalues are smaller than 10 eV\footnote{
The mass ranges given in this paragraph can be found in Ref.~\cite{FY}, and
in references therein.}.  The observed solar neutrino deficit may be explained
by MSW or vacuum $\nu_e$-$\nu_x$ oscillation, with $\Delta m^2 \sim
10^{-5}$~eV$^2$ or $\sim 10^{-10}$~eV$^2$, respectively, where $\Delta m^2$ 
is the difference in squared masses of the two relevant neutrino flavors.  
The atmospheric neutrino anomaly may result from $\nu_\mu$-$\nu_\tau$ 
oscillation with $\Delta m^2 \sim 10^{-2}$~eV$^2$.  If neutrinos 
are the hot dark matter in the Universe, then $m \sim 4$~eV is preferred.  
Furthermore, if neutrino masses are less than approximately 10~eV, 
it is possible to avoid a number of other cosmological constraints,
including those from big bang nucleosynthesis, overclosure of the
Universe, and the distortion of the cosmic microwave background radiation.
Finally, the recent LSND results suggest $\nu_e$-$\nu_\mu$ oscillation
with $\Delta m^2 \sim 10$~eV$^2$.  While it is unlikely that all the 
current neutrino anomalies will turn out to be real, it is still reasonable 
to conclude that the mass range below 10~eV is the most interesting one 
for neutrino physics\footnote{See Ref.~\cite{decaying} for counterexamples.}.

In implementing the seesaw mechanism, many have assumed that $M_{LR}$ 
should be comparable to the mass of the corresponding charged lepton or 
up-type quark, as a reasonable ansatz.  This is natural in many models, such 
as SO(10) grand unified theories where the Yukawa matrices of the charged 
leptons and/or the up-type quarks are related to those of the neutrinos by 
a gauge symmetry.  In models of this type, one needs 
$M_{RR} \simeq 10^{12}$~GeV to keep the tau neutrino mass below 10~eV, 
assuming $M_{LR}^{33} \sim m_{top}$.  While it is quite interesting that the 
the seesaw mechanism allows the light neutrino masses to serve as a probe
of physics at very high energy scales, it is unfortunate that the
mechanism cannot be proven directly in experiment.   In particular, there 
is no hope of producing such heavy right-handed neutrinos or studying 
their interactions at collider experiments in the imaginable future.

What we will demonstrate in this letter is that horizontal flavor
symmetries often naturally lead to a suppression of the matrix $M_{LR}$,
so that its eigenvalues are significantly {\em smaller} than the
corresponding charged lepton or up-type quark masses.  As is well known,
models with horizontal flavor symmetries have a separate motivation in
that they provide a natural framework for understanding the 
hierarchical form of other fermion mass matrices.  In theories where
the hierarchy between the electroweak scale and any fundamental high-energy 
scale (such as $M_{Planck}$) is stabilized by supersymmetry, horizontal 
symmetries have an additional virtue: they also restrict the form
of the soft supersymmetry-breaking scalar mass matrices and thereby may
suppress the large flavor changing neutral current processes expected 
when the superparticle spectrum is generic \cite{dlkns}.

What is significant about the suppression of $M_{LR}$ is that it 
allows us to construct models where the right-handed neutrino scale is 
at or slightly above the electroweak scale and where the light 
neutrino masses fall in a desirable range.  A right-handed neutrino 
scale near 1~TeV is very natural since we can imagine this situation
arising in a radiative breaking scenario: If the right-handed 
neutrinos $\nu$ transform under an additional U(1) gauge symmetry, 
then the mass squared of an exotic Higgs field $\rho$ that is also 
charged under this U(1) can be driven negative through renormalization 
group running, as a consequence of the Yukawa coupling $\rho \nu \nu$.  
This is completely analogous to the situation in the minimal 
supersymmetric standard model (MSSM), where the large top quark Yukawa 
coupling drives the up-type Higgs mass negative, triggering 
electroweak symmetry breaking at a scale comparable to the 
superparticle masses.  In the present case, the additional U(1) gauge 
symmetry is also spontaneously broken near the electroweak scale and 
the right-handed neutrinos develop Majorana masses.  Indeed, such additional 
U(1) gauge factors appear in many superstring 
compactifications \cite{extraU(1)}. The model we will present below 
demonstrates that the new $Z'$ boson is likely to be within the reach of 
the LHC, or the Tevatron after the main injector upgrade.

Below we will elaborate on these points by presenting a specific 
model based on the flavor group $(S_3)^3$.  Aside from bringing 
new physics down to the TeV scale, the neutrino physics of this
model is interesting in its own right.  Thus, this discussion 
complements the phenomenology presented in Ref.~\cite{carone1,carone2}
\footnote{The neutrino mass spectrum has been considered in the
context of other flavor groups \cite{hor}.  However, we know of
no reference in which a drastic suppression of the neutrino
Dirac mass matrix was either discussed or achieved.}.

\section{A Model}

The flavor group that we will assume is the discrete, non-Abelian 
gauge symmetry $(S_3)^3$.   While an Abelian horizontal symmetry
may be as effective in suppressing the Dirac mass matrix $M_{LR}$,
non-Abelian symmetries provide a natural means of suppressing
flavor-changing neutral current effects originating from superparticle 
loops, as we describe below.  This particular flavor symmetry has been 
discussed extensively in Refs.~\cite{carone1,carone2,hallmur}, so we will 
provide only a brief review.  

The group $S_3$ has three representations, ${\bf 2}$, ${\bf 
1}_S$ and ${\bf 1}_A$, where the latter two are trivial and nontrivial 
singlet representations, respectively.  The three generations of the 
quark chiral superfields $Q$, $U$, and $D$, are assigned to ${\bf 
2}+{\bf 1_A}$ representations of $S_3^Q$, $S_3^U$ and $S_3^D$, 
respectively.  The first two generation fields are embedded in a 
doublet to maintain the degeneracy of the corresponding squarks in the 
flavor symmetry limit.  After the flavor symmetry is spontaneously 
broken, the remaining approximate squark degeneracy is sufficient to 
suppress flavor changing neutral current effects from superparticle 
exchange, like those contributing to $K^0$-$\overline{K}^0$ mixing.  
The Higgs fields both transform as $({\bf 1}_A, {\bf 1}_A, {\bf 
1}_S)$'s, so that the top quark Yukawa coupling is invariant under the 
flavor symmetry group; this provides a natural explanation for the 
heaviness of the top quark relative to the other fermions.  The 
remaining Yukawa couplings can then be treated as small flavor 
symmetry breaking spurions, with the transformation properties
\begin{equation}
Y_U \sim \left(
\begin{array}{cc|c}
\multicolumn{2}{c|}{
({\bf \tilde{2}},{\bf \tilde{2}},{\bf 1}_S) }
& ({\bf \tilde{2}},{\bf 1}_S,{\bf 1}_S) \\ \hline
\multicolumn{2}{c|}{({\bf 1}_S,{\bf \tilde{2}},{\bf 1}_S)} &
({\bf 1}_S,{\bf 1}_S,{\bf 1}_S) \end{array} \right) \,\,\, ,
\end{equation}
\begin{equation}
Y_D \sim \left(\begin{array}{cc|c}
\multicolumn{2}{c|}{
({\bf \tilde{2}},{\bf 1}_A,{\bf 2})}
& ({\bf \tilde{2}},{\bf 1}_A,{\bf 1}_A) \\ \hline
\multicolumn{2}{c|}{({\bf 1}_S,{\bf 1}_A,{\bf 2})} &
({\bf 1}_S,{\bf 1}_A,{\bf 1}_A) \end{array}\right) \,\,\, ,
\label{eq:transp}
\end{equation}
where we use the notation ${\bf \tilde{2}} \equiv {\bf 2} \otimes {\bf
1}_A$,\footnote{${\bf \tilde{2}} = (a,b)$ is equivalent to ${\bf 2} =
(b, -a)$.}. Notice that the Yukawa matrices above involve at most 7 
irreducible multiplets of $(S_3)^3$.  In Ref.~\cite{hallmur}, $(S_3)^3$ was 
spontaneously broken by a set of `flavon' fields $\phi$, representing only 
four of these multiplets,
\begin{center}
\begin{tabular}{cc}
$\phi_1({\bf\tilde{2}}, {\bf\tilde{2}}, {\bf 1}_S)$  
&$\phi_2({\bf\tilde{2}},{\bf 1}_S, {\bf 1}_S)$ \\
$\phi_3({\bf \tilde{2}}, {\bf 1}_A, {\bf 2})$
& $\phi_4({\bf 1}_S, {\bf 1}_A, {\bf 1}_A)$ 
\end{tabular}
\end{center}
This was the minimal number needed to obtain realistic quark masses and 
mixing angles \cite{CM3}, assuming the Yukawa textures
\begin{equation}
Y_U = \left( \begin{array}{cc|c}
      h_u & h_c \lambda & - h_t V_{ub}\\
      {\cal O}(h_u)  & h_c & - h_t V_{cb} \\ \hline
      0 & 0 & h_t
            \end{array} \right) ,
\end{equation}
\begin{equation}
Y_D = \left( \begin{array}{cc|c}
      h_d & h_s \lambda & 0\\
       {\cal O}(h_d) & h_s & 0 \\ \hline 0 & 0 & h_b
            \end{array} \right).
\end{equation}
Here the $h_q$ are Yukawa couplings, and $\lambda \simeq 0.22$ is the
Cabbibo angle.  The hierarchical form of the Yukawa matrices presented 
above can be understood as a consequence of a sequential breaking of the 
flavor symmetry.  Since different components of a single multiplet will 
generally obtain comparable vevs at a given stage of symmetry breaking, we 
must assume that the $2\times 2$ blocks of $Y_U$ and $Y_D$ are each generated 
by two flavon fields that acquire vevs at somewhat different scales,
namely $\phi_1 = \phi_1' + \phi_1''$ and $\phi_2 = \phi_2' + \phi_2''$, 
where
\begin{equation}
\phi_1' = \left( \begin{array}{cc}
      0 & a' h_c \lambda \\ 0 & h_c
\end{array} \right) \,\,\, , \,\,\,
\phi_1'' = \left( \begin{array}{cc}
      h_u & {\cal O}(h_u) \\ {\cal O}(h_u)  & {\cal O}(h_u)
      \end{array} \right),
\end{equation}
and
\begin{equation}
\phi_2' = \left( \begin{array}{cc}
      0 & a h_s \lambda \\ 0 & h_s
\end{array} \right) \,\,\, , \,\,\,
\phi_2'' = \left( \begin{array}{cc}
      h_d & {\cal O}(h_d) \\ {\cal O}(h_d) & {\cal O}(h_d)
      \end{array} \right).
\end{equation}
Here $a$ and $a'$ are order one constants, with $a-a'=1$. 
Note that the entries labelled ${\cal O}(h_u)$ and ${\cal O}(h_d)$
above can be set to zero without noticeably affecting the quark
masses and mixing angles.  However, in any estimates where these entries 
are significant, we will assume they are nonvanishing with the magnitudes
given above.

The model is extended to the charged lepton sector by assigning the
$L$ and $E$ chiral superfields to ${\bf 2}+{\bf 1_A}$ representations
of $S_3^D$ and $S_3^Q$ respectively.  This is the only choice
that leads to a qualitative similarity between the down quark and charged 
lepton Yukawa matrices, while simultaneously forbidding
dangerous baryon-number-violating, Planck-suppressed operators (e.g.
$QQQL/M_{Pl}$) in the flavor symmetry limit.  Differences between
the charged lepton and down quark mass eigenvalues can originate
from fluctuations in the order $1$ coefficients that multiply the 
symmetry breaking operators; thus we assume that the electron-muon Yukawa 
matrix is given by $3 \phi_2'+ \frac{1}{3} \phi_2''$. 

If right-handed neutrinos $\nu$ are to be included in the model, then 
we must decide on their transformation properties under $(S_3)^3$.
A natural choice is to repeat the ${\bf 2}+{\bf 1_A}$ representation
structure of the other matter fields.  Aesthetics also suggests that we 
consider the possibility that the $\nu$ transform under $S_3^U$, the 
only $S_3$ factor that we haven't utilized in the lepton 
sector\footnote{In fact, the choice $S_3^Q$ would not give us the desired 
suppression of $M_{LR}$, which we would then expect to be of the same 
order as the charged lepton masses.  The choice $S_3^D$ does give us a 
high degree of suppression, but has other phenomenological difficulties, as 
we discuss in the Appendix.  It is worth noting that two of the three
natural charge assignments lead to a drastic suppression of $M_{LR}$.}.   
As before, we can then determine flavor 
structure of $M_{LR}$ and $M_{RR}$:
\begin{equation}
M_{RR} \sim \left(
\begin{array}{cc|c}
\multicolumn{2}{c|}{
({\bf 1}_S,{\bf 2}+{\bf 1}_S,{\bf 1}_S) }
& ({\bf 1}_S,{\bf \tilde{2}},{\bf 1}_S) \\ \hline
\multicolumn{2}{c|}{({\bf 1}_S,{\bf \tilde{2}},{\bf 1}_S)} &
({\bf 1}_S,{\bf 1}_S,{\bf 1}_S) \end{array} \right)
\end{equation}
\begin{equation}
M_{LR} \sim \left(\begin{array}{cc|c}
\multicolumn{2}{c|}{
({\bf 1}_A,{\bf \tilde{2}},{\bf 2})}
& ({\bf 1}_A,{\bf 1}_S,{\bf 2}) \\ \hline
\multicolumn{2}{c|}{({\bf 1}_A,{\bf \tilde{2}},{\bf 1}_A)} &
({\bf 1}_A,{\bf 1}_S,{\bf 1}_A) \end{array}\right)
\end{equation}
Given our assumption that flavor symmetry breaking originates only
from the vevs of the four flavons $\phi_i$, it is possible to construct 
the $(S_3)^3$ representations shown above from products of the flavons.
Thus, we will obtain the entries of the neutrino mass matrices in terms
of products of the quark Yukawa couplings.  A useful way to display
our result is to express the Yukawa couplings and the third generation 
mixing angles in terms of powers of the Cabibbo angle: $h_u\sim \lambda^8$, 
$h_c\sim \lambda^4$, $h_t \sim 1$, $h_d \sim \lambda^7$, $h_s \sim \lambda^5$,
$h_b\sim\lambda^3$, $V_{cb}\sim\lambda^2$, and $V_{ub}\sim \lambda^3$.
We then  obtain:
\begin{equation}
M_{RR}\sim c_0 \langle \rho \rangle \left(\begin{array}{ccc}
1+c_1 \lambda^6 & c_1  \lambda^{10} & c_2 \lambda^{10} \\
c_1  \lambda^{10} & 1 - c_1 \lambda^6 & c_2 \lambda^6 \\
c_2 \lambda^{10}  &  c_2 \lambda^6    & c_3 \end{array}\right)
\label{eq:mrr}
\end{equation}
\begin{equation}
M_{LR}\sim \langle H_u \rangle \lambda^{10} \left(\begin{array}{ccc}
d_1 \lambda & d_1 \lambda^5  & d_3 \lambda \\
-d_1        & d_1 \lambda^3 & -d_3 \lambda^2 \\
d_2 & -d_2 \lambda^3 & d_4 \lambda^2 \end{array}\right)
\label{eq:mlr}
\end{equation}
where $\rho$ is the field whose vacuum expectation value
determines the right-handed neutrino mass scale, and the
$c_i$ and $d_i$ are order $1$ coefficients.  Notice that $M_{LR}$ is 
suppressed by an overall factor of $\lambda^{10}$.  This suppression has 
a simple interpretation.  Had we chosen $\nu$ to transform as 
a ${\bf 2}+{\bf 1}_A$ under $S_3^Q$, we would expect flavor symmetry 
breaking to occur at the same order in the symmetry breaking as the
down quark Yukawa matrix.  However, with $\nu$ transforming
under $S_3^U$, the matrix $M_{LR}$ has a flavor symmetry structure
that is completely different from all the other Yukawa matrices.
Since the quark Yukawa couplings are the origin of flavor symmetry
breaking in this model, we can only construct $M_{LR}$ by going to 
{\em one higher order} in the symmetry breaking parameters, which, roughly 
speaking, is of the order of a typical Yukawa coupling squared.  The
precise value of the suppression is specific to the group theory of the 
given flavor model.

We may now compute the quantity of interest, the Majorana mass
matrix for the light neutrino states.  In the case
of three flavors, the seesaw mechanism described earlier can
be generalized:
\begin{equation}
M_{LL} = M_{LR} M_{RR}^{-1} (M_{LR})^T \,\,\, .
\end{equation}
From Eqs.~(\ref{eq:mrr}) and (\ref{eq:mlr}) above, we obtain
{\samepage
\begin{eqnarray}
\lefteqn{M_{LL} = \frac{\langle H_u \rangle^{2} \lambda^{20}}{c_0 \langle \rho
\rangle} \times} & & \nonumber \\
&
\left(\begin{array}{ccc} 
(d_1^2 + d_3^2/c_3)\lambda^2 
& -d_1^2 \lambda - (d_3^2/c_3) \lambda^3
& d_1 d_2 \lambda + (d_3 d_4/c_3)\lambda^3 \\
-d_1^2 \lambda - (d_3^2/c_3) \lambda^3   
&  d_1^2 + (d_3^2/c_3) \lambda^4  
&  -d_1 d_2 - (d_3 d_4/c_3) \lambda^4 \\
d_1 d_2 \lambda + (d_3 d_4/c_3)\lambda^3    
&  -d_1 d_2 - (d_3 d_4/c_3) \lambda^4 
& d_2^2 + (d_4^2/c_3) \lambda^4  \end{array} \right)
& \nonumber \\
\label{eq:mllr}
\end{eqnarray}}
where we have retained higher order terms that lift a zero eigenvalue 
present at leading order.  Notice that if $\langle \rho
\rangle \approx 1$ TeV, $\lambda \approx 0.22$ and $\langle H_u \rangle
\approx 175$ GeV, the overall scale of this matrix is of order $2 \times
10^{-3}$ eV.  This is approximately the correct magnitude to obtain the MSW
solution to the solar neutrino problem, as we will see below. Given the 
very high power in $\lambda$, the predicted mass range can easily vary 
within an order of magnitude.
 
Perhaps the easiest way to study the physical implications of 
Eq.~(\ref{eq:mllr}) is to rotate to a new basis $\nu_L' = U \nu_L$, 
where
\begin{equation}
U = 
\left(\begin{array}{ccc}  1 & 0 & 0 \\ 
0 & -d_1/n & d_2/n \\
0 &  d_2/n & d_1/n \\ \end{array}\right)
\end{equation}
with $n=(d_1^2+d_2^2)^{1/2}$. Then $M_{LL}$ becomes
{\samepage
\begin{eqnarray}
\lefteqn{M'_{LL} = \frac{\langle H_u \rangle^{2} \lambda^{20}}{c_0 \langle
\rho
\rangle} \times } & \nonumber \\
& 
\left(\begin{array}{ccc} 
(d_1^2 + d_3^2/c_3)\lambda^2 & d_1 n \lambda &
(d_3/c_3)(d_1 d_4 - d_2 d_3)\lambda^3/n \\
d_1 n \lambda    & n^2  & {\cal O}(\lambda^4) \\
(d_3/c_3)(d_1 d_4 - d_2 d_3)\lambda^3/n & {\cal O}(\lambda^4) & 
{\cal O}(\lambda^4)
\end{array} \right) .
& \nonumber \\
\label{eq:mllp}
\end{eqnarray}}
We have retained order one coefficients only where they are relevant 
to our estimates below.  Note that we principally will be interested
in the $1$-$2$ mixing, since $\Delta m^2_{12}$ and $\sin^2 2 \theta_{12}$
are in the appropriate range for the the MSW solution.  However, we will 
check that the $1$-$3$ mixing (which corresponds to a much smaller 
$\Delta m^2_{13}$) does not lead to an unacceptable depletion of electron 
neutrinos that would be observable in terrestrial solar neutrino 
experiments\footnote{The $\nu_\mu$-$\nu_\tau$ mixing is not observable 
experimentally, given the small value of $\Delta m^2$.}. 

Let us consider the neutrino mixing quantitatively. The $1$-$2$ mixing 
angle from Eq.~(\ref{eq:mllp}) is given by
\begin{equation}
      \sin^{2} 2\theta_{12} = \frac{4 d_{1}^{2} (d_1^2 + d_2^2) \lambda^{2}}
      {(d_{1}^{2} + d_{2}^{2}-(d_1^2 +d_3^2/c_3)\lambda^2)^{2} 
         + 4 d_{1}^{2} (d_1^2 + d_2^2) \lambda^{2}} .
      \label{mixing}
\end{equation}
If we choose the order 1 coefficients $d_1=0.4$, $d_2=2$,
$d_3=1$, $c_0=2$, $c_3=1$, and we set  $\langle \rho \rangle =2$~TeV,
we find
\begin{eqnarray}
\sin^2 2\theta_{12} & = & 7.6 \times 10^{-3} \,\,\, , \\
\Delta m^2_{12} & = & 5.1 \times 10^{-6} \mbox{ eV}^2 \,\,\, .
\end {eqnarray}
This is consistent with the preferred range of the small angle MSW 
solution to the solar neutrino problem \cite{Bahcall-fit}
\begin{eqnarray}
\sin^2 2\theta & = & 3 \times 10^{-3} \mbox{--} 1.1 \times 10^{-2} 
\,\,\, , \\
\Delta m^2 & = & 3 \times 10^{-6} \mbox{--} 1 \times 10^{-5} \mbox{ eV}^2
\,\,\, .
\end{eqnarray}
Note that all the order 1 coefficients in this example are within the 
range $0.4$--$2$, and the solution involved no fine-tuning.  For
this parameter set, $\sin^2 2\theta_{13} \sim 0.1$ and 
$\Delta m^2_{13} = 7 \times 10^{-10}$~eV$^2$.  With 
$\Delta m^2_{13}$ this small, the $1$-$3$ mixing does not lead 
to a significant depletion of the electron neutrino flux, and
can be ignored.

It is also possible to achieve the large angle MSW solution, though
in this case some fine-tuning is involved.   If we choose $d_1=1.1$,
$d_2=0.4$, $d_3=2.5$, $c_0=2/3$, $c_3=0.4$ and $\langle \rho \rangle =1$~TeV,
we obtain
\begin{eqnarray}
\sin^2 2\theta_{12} & = & 0.5 \,\,\, ,\\
\Delta m^2_{12} & = & 1.8 \times 10^{-5} \mbox{ eV}^2 \,\,\, ,
\label{eq:lar}
\end {eqnarray}
which is consistent with the preferred range for the large angle MSW 
solution \cite{Bahcall-fit}
\begin{eqnarray}
\sin^{2} 2\theta & = & 0.5 \mbox{--} 0.9 \,\,\, , \\
\Delta m^{2} & = & (1\times 10^{-5} \mbox{--} 1\times 10^{-4}) 
\mbox{ eV}^{2}
\,\,\, . \label{large-angle}
\end{eqnarray}
In this parameter set, the order 1 coefficients fall within the 
range $0.4$--$2.5$.  Unfortunately, $\Delta m^2_{13} = 5 \times
10^{-6}$ implies that there would be significant depletion of
electron neutrinos observed in the $^{71}$Ga experiments, unless 
$\sin^2 2\theta_{13} \lesssim 10^{-4}$ \cite{FY}.   This can be
achieved in our model, providing we tolerate a 7\% fine-tuning, 
$d_1 d_4-d_2 d_3 \sim 0.07$.

Thus, it seems that the small angle MSW solution arises more naturally
in our model.  This is encouraging given that the small angle solution 
provides a better fit to the current data than the large angle 
one \cite{Bahcall-fit}.  It is worth pointing out that the superKamiokande 
and SNO experiments are likely to see a distortion in the electron 
energy spectrum if the small angle solution is correct. Such a distortion 
would be an unambiguous indication of neutrino oscillation since it does 
not rely on the normalization of the solar neutrino flux, which is extremely 
sensitive to the core temperature of the Sun, scaling as $\sim T^{18}$.  

\section{Right-handed Neutrino Scale}

In the example above, we saw that a completely reasonable theory of neutrino 
masses and mixings could be obtained with neutrinos in the $10^{-3}$~eV 
range, even when the right-handed scale $\langle \rho \rangle \approx 1$ TeV.
In this section, we show how $\langle \rho \rangle$ can naturally 
be slightly larger than the electroweak scale in a radiative breaking 
scenario.  A new gauge boson becomes massive when the $\rho$ field
acquires a vev, and could easily lie just beyond the current limits 
set from direct searches at the Tevatron.

We will assume an additional U(1) gauge symmetry, under which the 
right-handed neutrinos have charge $+1$.  The simplest choice for the 
purpose of illustration is U(1)$_{B-L}$, since only right-handed neutrinos 
are required to render this symmetry nonanomalous in the MSSM.  In 
addition, this extension of the MSSM preserves unification of the ordinary 
gauge coupling constants\footnote{The particular flavor symmetry group, 
$(S_{3})^{3}$, is difficult to implement in a conventional grand
unified model.  However, string unification could be a viable
option in this scenario.}.  We also assume the presence of a pair of exotic 
Higgs fields with equal and opposite B-L charges, $\rho_{+2}$ and $\rho_{-2}$.
The superpotential that we will consider is
\begin{equation}
W = \alpha S \rho_{+2} \rho_{-2} - \frac{\beta}{3} S^3 
+ \gamma S H_u H_d + \frac{1}{2} Y_{RR} \rho_{-2} \nu \nu + Y_{LR} L H_u \nu
\label{eq:suppot}
\end{equation}
where $S$ is a gauge singlet, and $\alpha$, $\beta$ and $\gamma$
are coupling constants.  The matrices $M_{RR}$ and $M_{LR}$ described earlier
correspond to $Y_{RR} \langle \rho_{-2} \rangle$ and 
$Y_{LR} \langle H_u \rangle$, respectively.

The potential of this theory is given by
\begin{eqnarray}
V &=& |\alpha \rho_{+2} \rho_{-2} - \beta S^2|^2 + |\alpha S \rho_{-2}|^2
+ |\alpha S \rho_{+2}|^2 + 2 g^2 (|\rho_{+2}|^2 - |\rho_{-2}|^2)^2
\nonumber\\
&&
+ m_+^2 |\rho_{+2}|^2 + m_-^2 |\rho_{-2}|^2 + m_s^2 |S|^2
-(A_\alpha \alpha S \rho_{+2} \rho_{-2} - \frac{1}{3} A_\beta \beta S^3 
+ \mbox{h.c})\nonumber \\
\label{eq:thepot}
\end{eqnarray}
where $g$ is the U(1)$_{B-L}$ gauge coupling, and  
($m_+^2$, $m_-^2$, $m_s^2$, $A_\alpha$, $A_\beta$) are soft supersymmetry 
breaking masses and trilinear couplings.  In a radiative breaking scenario, 
the Yukawa couplings $Y_{RR}$ drive the soft squared mass $m_-^2$ negative in
the renormalization group running, so we will look for a minimum of
this potential assuming that $m_-^2 < 0$, with the remaining mass
squared parameters positive.  For $\alpha=\beta=\gamma=0.1$ (to be
explained below), $g=0.3$, and the following dimensionful input 
parameters (in units of GeV)
\begin{center}
\begin{tabular}{ccccc} 
$A_\alpha$ & $A_\beta$ & $m_{+}^{2}$ & $m_{-}^{2}$ & $m_{S}^{2}$ 
\\ \hline
$200$  & $100$   &$(256)^2$  & $-(274)^2$ & $(191)^2$ \\
\end{tabular}
\end{center}
we find the vacuum expectation values
\begin{center}
\begin{tabular}{ccc} 
$\langle \rho_{+2} \rangle$ & $\langle \rho_{-2} \rangle$ 
& $\langle S \rangle$ 
\\ \hline
$1.3$ TeV & $1.4$ TeV & $0.7$ TeV \\
\end{tabular}
\end{center}
and the mass squared eigenvalues
\begin{center}
\begin{tabular}{ccc}
$M^{s}_1$ & $M^{s}_2 $ & $M^{s}_3$ \\ \hline
$(122 \mbox{ GeV})^2$   & $(322 \mbox{ GeV})^2$ 
& $(1655 \mbox{ GeV})^2$ \\
$M^{p}_1$ & $M^{p}_2$ & $M^{p}_3$ \\ \hline 
$(191 \mbox{ GeV})^2$ & $(332 \mbox{ GeV})^2$ & $0$ \\
\end{tabular}
\end{center}
where the $0$ eigenvalue corresponds to the degree of freedom
that is `eaten' by the U(1) gauge boson.  Note that this solution 
has the desired feature that $m_-^2 < 0$ while $m_+^2$ and $m_S^2$ 
remain positive.  The positivity of the scalar ($s$) and pseudoscalar ($p$) 
mass eigenvalues indicates that we have found a local minimum of the 
potential, which will suffice for our purposes.  Note that 
$\langle \rho_{-2} \rangle$ is at 2 TeV, giving us the desired right-handed 
neutrino scale.  Furthermore, with $\langle S \rangle$ at 1 TeV, and the
auxiliary component $F_s = \alpha \rho_{+2} \rho_{-2} - \beta S^2$ at
(374~GeV)$^2$, we generate a $\mu$ parameter of order $100$ GeV,
and $B\mu$ of order (118~GeV)$^2$; this is in the proper range
for electroweak symmetry breaking.   Finally, we see that
the $Z'$ gauge boson develops a mass of $1.7$~TeV, which is
within the expected reach of the LHC, as we describe below.

The choice of couplings $\alpha=\beta=\gamma=0.1$ was
convenient for the purposes of illustration, since it allowed
for a separation between the B-L and electroweak scales, so
that the Higgs doublet vevs could be neglected in the minimization of
Eq.~(\ref{eq:thepot}).  We could have chosen these parameters 
to be of order one if we had allowed smaller $\rho$ vacuum 
expectation values, and analysed the full potential without approximation.  
This is beyond the scope of the present work.  However, it is worth 
pointing out that smaller $\rho$ vevs are not immediately excluded by 
the current bounds on the $Z'$ boson mass.  With our choice $g=0.3$
(which in the given normalization is the approximate value one would
expect from gauge coupling unification), the strongest constraint
on the $Z'$ mass comes from direct collider searches in the dilepton 
channel, yielding $m_{Z'} < 425$ GeV \cite{pdg}; this implies 
$\langle \rho \rangle > 500$~GeV, assuming equal vevs for
$\rho_{+2}$ and $\rho_{-2}$.  Thus, it seems there is no reason why
the $Z'$ in our model could not have a mass that is within the discovery 
reach of the Tevatron after the main injector upgrade, 
$m_{Z'}\lesssim 800$~GeV \cite{tev}.  The corresponding reach at
the LHC, $m_{Z'}\lesssim 4.3$~TeV \cite{tev}, suggests that our model
would also be testable in the limit 
$\langle \rho \rangle \gg \langle H \rangle$, that we obtained earlier
by chooising $\alpha=\beta=\gamma=0.1$.  Such a parameter set can be 
justified by a symmetry argument, so it is not necessarily just 
a convenient limit.  For example, we may imagine extending the flavor 
symmetry to the nonanomalous, discrete gauge group $(S_3)^3 \times Z_2$, 
where the $Z_2$ factor acts on the field $S$ which is odd.  In addition, we 
may assume there is a new flavon field $\zeta$ that is also $Z_2$ odd, and 
$\langle \zeta \rangle/M_{PL} \approx 1/10$. A spurion analysis similar 
to the one presented earlier would then suggest that all couplings involving 
an odd number of $S$ fields should naturally be suppressed by $1/10$. This
would yield an effective theory in which $\alpha$, $\beta$ and $\gamma$ 
are all of the same order, and are all relatively small.  

In the scenario that we have described, there are two other issues
that deserve comment.  First, we have not justified the purely trilinear 
form of the superpotential in Eq.~(\ref{eq:suppot}).  Second, if B-L 
is gauged down to low energies, it is not immediately clear how we 
can generate and preserve a cosmic baryon asymmetry, taking into account 
the electroweak sphaleron effect\footnote{Note that we could have chosen a 
different gauged U(1) group, for example from E$_6$ $\rightarrow$ 
SO(10) $\times$ U(1) breaking, so that the right-handed neutrinos are 
charged but $B-L$ is not a gauge symmetry.  In this case, the constraints 
from baryogenesis are significantly weaker.}.  The answers to these questions 
are somewhat more speculative than the main points of this letter, but we 
include them for completeness.   On the first issue, one might imagine the 
relevant fields are simply a part of the massless spectrum of string theory 
and the superpotential in Eq.~(\ref{eq:suppot}) gives the complete set of
renormalizable interactions. Alternatively, we could impose a 
global $Z_3$ or $R$-symmetry that restricts all renormalizable interactions 
to be trilinear in form.  However, since global symmetries are thought to be
violated by Planck-scale effects, this symmetry may not apply to the
Planck-suppressed operators from which flavor symmetry breaking
originates.  On the second issue, we may imagine that baryogenesis proceeds 
via the Affleck-Dine mechanism \cite{AD} with large enough baryon to entropy 
ratio $n_B/s$ so that the Bose condensate of the squarks and sleptons make 
the sphalerons heavy enough to preserve the $B$ and $L$ asymmetries even 
when $B-L=0$ \cite{DMO}.  For instance, consider the operator
\begin{equation}
\frac{h_c}{M^2_*} Q_1 \overline{U}_1^* L_2 \overline{D}^*_2
\end{equation}
where the charm Yukawa coupling originates from the usual spurion 
analysis, $M_*$ is the reduced Planck mass, and the subscript indicates
the generation.  This operator respects all the unbroken symmetries in the 
low-energy effective theory, including U(1)$_{B-L}$.  The estimate of 
the final baryon-to-entropy ratio that it generates is given by
\begin{equation}
\frac{n_B}{s} \sim g_*^{-1/4} h_c \left(\frac{M_*}{m}\right)^{1/2}
      \left(\frac{\phi_0}{M_*}\right)^4,
\end{equation}
where $m$ is a typical supersymmetry-breaking scalar mass, $g_*$ 
is the effective number of degrees of freedom in the Early Universe, and 
$\phi_0$ is the initial amplitude of the scalar fields, when they begin 
to oscillate.  In order to have sufficient suppression of the sphaleron 
effects, we need $n_B/s \gtrsim 10^{-2}$ which translates into the 
bound $\phi_0 \gtrsim 10^{16}$~GeV.  Such a large initial amplitude
may be generated if there is a scalar field with a negative mass
squared during inflation, in a model with a non-minimal K\"ahler 
potential coupling to the inflaton \cite{DRT} or in no-scale 
supergravity \cite{GMO}.  Subsequent entropy production is then needed 
to dilute $n_B/s$ down to the value required by nucleosynthesis; this
possibly may be provided by the decay of a Polonyi-like field or by a 
late inflation with an $e$-folding of order 7, like that in Ref.~\cite{LS}.

\section{Conclusions}

We have shown that flavor symmetries allow for a class of models
in which neutrino masses can be made extremely small, without 
requiring the mass scale of right-handed neutrinos to be larger than 
the electroweak scale.  While our model was based on a discrete, 
non-Abelian family symmetry, the general idea should also be
applicable to a large number of other flavor models, in particular
those involving Abelian horizontal symmetries.  The scenario we have outlined 
supports the notion that there are a wide class of models where right-handed 
neutrino Yukawa couplings may lead to a radiative breaking of 
additional U(1) gauge groups, leading naturally to $Z'$ bosons with 
masses less than a few TeV.  With the LHC expected to discover a $Z'$ 
bosons with standard model couplings up to about 5 TeV \cite{tev}, this 
scenario can be tested definitively in the future.

\begin{center}               
{\bf Acknowledgments} 
\end{center}
This work was supported in part by the Director, Office of 
Energy Research, Office of High Energy and Nuclear Physics, Division of 
High Energy Physics of the U.S. Department of Energy under Contract 
DE-AC03-76SF00098.  HM was also supported in part by the National 
Science Foundation under grant PHY-95-14797, and by the Alfred 
P. Sloan Foundation.

\appendix
\section{Another Model}
\setcounter{equation}{0}

For completeness, we present the mass matrix $M_{LL}$ that we obtain
when the right-handed neutrinos $\nu$ transform as a ${\bf 2}+{\bf 1}_A$ 
under $S_3^D$ rather than $S_{3}^{U}$.  We obtain
\begin{equation}
M_{LL} \sim \frac{\langle H_u \rangle^2 \lambda^{16}}{c_0\langle \rho \rangle}
\left(\begin{array}{ccc}
d_1^2 + d_3^2/c_3 & -d_3^2/c_3 \lambda & -2d_1 d_2 \lambda \\
-d_3^2/c_3 \lambda &  d_1^2 &  d_1 d_2 \\
-2d_1 d_2 \lambda & d_1 d_2 & d_2^2 \end{array} \right) \,\,\, ,
\label{eq:newres}
\end{equation}
where the coefficients $c_i$ and $d_i$ multiply the same entries of
$M_{RR}$ and $M_{LR}$ as in the example presented in the text.  In this
case, if we choose $\langle \rho \rangle \approx 1$ TeV, and $\langle H_u
\rangle \approx 175$ GeV, the overall scale of this matrix is of
order $1$~eV.  Again, the flavor symmetry structure of the model is 
responsible for a large suppression of the Dirac mass matrix, of
order $\lambda^8$.

Notice from the determinant of Eq.~(\ref{eq:newres}) that two of 
the eigenvalues are of order 1~eV, while the remaining one is 
roughly $\lambda^2$ smaller.  Thus, the corresponding $\Delta m^2$ values 
do not suggest a natural solution to either the solar, or atmospheric 
neutrino problems.  The $1$~eV mass scale is appropriate for the 
neutrinos to be hot dark matter candidates.  However, with masses in 
this range the large mixing between the second and third generations 
is problematic in light of the bounds from disappearance experiments 
that search for $\nu_\mu \rightarrow \nu_x$.  These searches completely 
exclude $\sin^2 2\theta_{23} > 0.1$ for $\Delta m^2 = 1$~eV$^2$, and place
significantly stronger bounds for $\Delta m^2$ between 1--100~eV$^2$
\cite{gruwe}.  Thus, the hot dark matter solution seems possible 
only for very small $d_1/d_2< 0.15$ with $d_2$ of order 1, implying
a rather unnatural value for $d_1$. For a somewhat smaller overall scale, 
(choosing $c_0=2$ and $\langle \rho \rangle=3$ TeV, for example), 
we can obtain an acceptable theory, which explains no neutrino
anomalies, but implies $\nu_\mu$-$\nu_\tau$ mixing at a level that is 
likely to be measurable at the CHORUS or NOMAD experiments.

\end{document}